\documentstyle[12pt]{article}
\def\s{\psi}
\def\s{\psi}
\def\bs{\bar\psi}
\def\a{\alpha}
\def\b{\beta}
\def\g{\gamma}

\def\d{\delta}
\def\e{\epsilon}
\def\t{\theta}
\def\be{\begin{equation}}
\def\ee{\end{equation}}
\def\p{\partial}
\def\ber{\begin{eqnarray}}
\def\eer{\end{eqnarray}}
\begin{document}

\begin{center}
{\Large\bf Anomalies in noncommutative gauge theories, Seiberg-Witten transformation  and Ramond-Ramond couplings }
\vskip 1 true cm
{\bf Rabin Banerjee}\footnote{On leave from S.N.Bose Natl. Ctr. for Basic Sciences,
Calcutta, India; e-mail:rabin@post.kek.jp;  rabin@bose.res.in}
\vskip .8 true cm
Institute of particle and nuclear studies,

High Energy Accelerator Research Organisation (KEK),

  Tsukuba 305-0801, Japan
\end{center}

\bigskip

\centerline{\large\bf Abstract}
\medskip

We propose an exact expression for the unintegrated form of the star gauge invariant axial anomaly in an arbitrary even dimensional noncommutative gauge theory.  The proposal is based on the inverse
Seiberg-Witten map and identities related to it, obtained earlier by comparing Ramond-Ramond couplings in different  descriptions. The integrated anomalies are expressed in terms of a simplified version of the Elliott formula involving the noncommutative Chern character. These anomalies, under the Seiberg-Witten transformation, reduce to the ordinary axial anomalies. Compatibility with existing results of anomalies in noncommutative theories is established.

\newpage
\section{\bf{Introduction}}

The subject of anomalies has a long history, playing a fundamental role in the development of modern gauge theories \cite{b}. Of no less consequence is its role in noncommutative field theories, both from mathematical and physical standpoints. Anomalies in this context have been widely discussed \cite{as}-\cite{n} recently ,  including their possible implications in model building on noncommutative spaces \cite{w}. 

The gauge invariant axial $U(1)$ anomaly in ordinary QED, historically the first studied divergence anomaly in field theory, led to far reaching consequences. The puzzle of the decay process $\pi^0\rightarrow 2\gamma$, which was observed in nature but theoretically forbidden,  was resolved by this anomaly, to a remarkable accuracy. There are also other types of anomalies that are associated with the gauge currents of the theory and their structure is important for obtaining the anomaly canceling conditions in order to make such theories meaningful. However the structure of the axial anomaly influences  these types of anomalies also. In nonabelian chiral gauge theories, the covariant anomaly has a functionally similar form as the axial anomaly. 
Knowing the covariant anomaly, it is possible to compute the gauge variation of the effective action, thereby obtaining the consistent anomaly. This approach, developed in \cite{rb}, is opposite to the usual cohomological approach, where the consistency condition is first used to obtain the consistent anomaly, from which the covariant anomaly is calculated \cite{b}.

Anomalies in noncommutative gauge theories (i.e. gauge theories defined on noncommutative space) seem to provide a richer structure than their conventional counterparts. It is basically linked with the noncommutativity of the star product that is used in the definition of composite objects. By reversing the order of the fermions appearing in the definition of the current, it is possible to get star gauge invariant and star gauge covariant currents even in the $U(1)$ theory. The computation of the covariant anomaly has been done by several methods, but the same is not true for the invariant anomaly. Indeed, calculations for this anomaly exist only in their integrated versions, so that the unintegrated structure for the invariant anomaly is not known. The result for the unintegrated anomaly, up to the first order only in the noncommutative parameter, was provided by us in \cite{bg}.

Recalling the fundamental role played by the gauge invariant $U(1)$ anomaly in the usual case, we feel that a proper evaluation of the unintegrated star gauge invariant anomaly is crucial for a better understanding of anomalies in noncommutative gauge theories. Indeed, continuing the parallelism with the usual case, the structure of this anomaly could  yield the covariant anomaly.
Another aspect which has hardly received any attention in this context is the significance of the Seiberg-Witten transformation \cite{sw}.  Since this transformation maps the noncommutative objects with their commutative equivalents, it becomes justifiable to ask the role, if any, of this map in relating the anomalies in the different descriptions.

In this paper we propose an exact expression for the 2n-dimensional unintegrated star gauge invariant anomaly.
This proposal is based on the (inverse) Seiberg-Witten map and a series of identities related to it.
Both the inverse solution  (conjectured in \cite{hl} and later proved in \cite{lm, oo, ms}) and the identities \cite{lm} were found by comparing, in different descriptions, the couplings of D-branes to Ramond-Ramond potentials in a constant NS-NS B-field.  The integrated version of these anomalies are equivalent to the (integrated) covariant anomalies. They are expressed by a simplified form of the Elliott formula \cite{e} involving the noncommutative Chern character. The Seiberg-Witten transformation maps this formula to the usual formula involving the ordinary Chern character, from which the usual $U(1)$ anomalies are obtained. We also comment on an extension to include chiral anomalies. Compatibility with existing results is established.

This paper is organised as follows: after putting the problem in a proper perspective, which also helps in setting up the notations, we discuss our proposal for the anomaly in section 2; section 3 gives an alternative derivation, highlighting the connection of the simplified Elliott formula with the integrated anomalies; section 4 contains some discussions.

We consider a noncommutative $U(1)$ gauge theory with a massless Dirac fermion transforming in the fundamental representation, whose action is given by,
\be
S=i\int d^{2n}x (\bs*\gamma^\mu D_\mu\s) ;\,\,\, D_\mu\s=\p_\mu\s +iA_\mu*\s
\label{0}
\ee

Contrary to the usual (commutative) case, the axial current in such a gauge theory has two distinct structures given by,
\be
a)\,\,\,  j^{\mu(5)}=\bs *\g^\mu\g^5\s, \,\,\,  b)\,\,\,  J^{\mu(5)}=-\s^t *(\g^\mu\g^5)^t\bs^t
\label{1}
\ee
The difference arises because the star product, as defined below, is noncommutative,
\be
f(x)*g(x)=
e^{-\frac{i}{2}\t_{\mu\nu}{\p\over{\p\eta_\mu}}{\p\over{\p\d_\nu}}}f(x+\eta)g(x+\d)|_{\eta=\delta=0}
\label{2}
\ee
where $\t_{\mu\nu}$ is a real constant antisymmetric matrix manifesting the noncommutativity of the coordinates,
\be
[x_\mu, x_\nu]=-i\t_{\mu\nu}
\label{3}
\ee

It is trivial to note that the definitions (a) and (b) are equivalent in the usual theory. The simple reordering of the variables has quite nontrivial consequences. Under the local gauge transformations,
\be
\s(x)\rightarrow \s(x)'=U(x)*\s(x); \bs(x)\rightarrow\bs(x)'=\bs(x)*U^{-1}(x)
\label{4}
\ee
where,
\be
U(x)=*exp(i\alpha(x))=1+i\alpha(x)-{1\over 2}\alpha(x)*\alpha(x)+O(*\alpha^3)
\label{5}
\ee
the current (a) remains invariant (because $U^{-1}(x)*U(x)=1$), but the current (b) transforms covariantly; i.e.,
\be
j^{'\mu(5)}(x)=j^{\mu(5)}(x)
\label{6}
\ee

\be
J^{'\mu(5)}(x)=U(x) * J^{\mu(5)}(x) * U^{-1}(x)
\label{7}
\ee
Correspondingly, the classical equations of motion lead to the conservation and covariant conservation, respectively, of these currents,
\be
a)\,\,\,  \p_\mu j^{\mu(5)}=0;\,\,\, b)\,\,\,  D_\mu J^{\mu(5)}=\p_\mu J^{\mu(5)}+i[A_\mu, J^{\mu(5)}]_*=0
\label{8}
\ee
where $[A, B]_*=A*B - B*A$.  The classical result gets modified by quantum corrections, as happens for the usual case. It has been found that the covariant current leads to the covariant anomaly,
\be
D_\mu J^{\mu(5)}=-{1\over{16\pi^2}}\e^{\mu\nu\rho\sigma}F_{\mu\nu} * F_{\rho\sigma}=
-{1\over{16\pi^2}} (*)F\wedge F
\label{9}
\ee
where the wedge notation has been introduced and the field tensor,
\be
F_{\mu\nu}=-i[D_\mu, D_\nu]_*=\p_\mu A_\nu-\p_\nu A_\mu +i[A_\mu, A_\nu]_*
\label{10}
\ee
transforms covariantly under the gauge transformation,
\be
F_{\mu\nu}\rightarrow U* F_{\mu\nu} *U^{-1}
\label{11}
\ee
The one form potential, on the other hand, transforms as,
\be
A_\mu\rightarrow U*A_\mu * U^{-1} - i U*\p_\mu U^{-1}
\label{a}
\ee
The star gauge covariant anomaly (\ref{9}) is the noncommutative generalisation of the ordinary gauge invariant axial anomaly in four space-time dimensions,
\be
{\cal A}=- {1 \over{16\pi^2}}\e^{\mu\nu\rho\sigma}f_{\mu\nu} f_{\rho\sigma}
\label{12}
\ee
where $f_{\mu\nu}$ is the ordinary gauge invariant field tensor,
\be
f_{\mu\nu}= \p_\mu a_\nu-\p_\nu a_\mu
\label{13}
\ee

 A similar generalisation is expected to be valid for arbitrary even dimensions and there are calculations which support this viewpoint.

The situation is however obscure for the anomaly of the star gauge invariant current. This is because, in contrast to $F_{\mu\nu}$ transforming covariantly, there is no simple object that remains invariant under the star gauge transformations. 
Before proceeding, it is useful to recapitulate the present status. Initially it was felt that there was no gauge invariant anomaly. This was suggested \cite{ cpm, bs, ik}, on the basis of diagrammatic analysis. The point is that the phase of the exponential in the definition of the star product involves derivatives which, in a loop calculation, may or may not convert to the loop momenta. If this phase does not depend on the loop momenta, it can be taken outside the integral. The result is therefore, apart from an overall phase, identical to the result in the ordinary case. Such diagrams are called planar diagrams which can be easily evaluated. If the phase depends on the loop momenta, then it cannot be pushed outside the integral and the resulting graphs are called non-planar graphs. Such graphs are responsible for the anomaly in the star gauge invariant anomaly. Non-planar one loop graphs were believed to be UV finite thereby leading to the conclusion of a vanishing anomaly. However an important subtlety was pointed out \cite{as, bg, alt, n}. Although these graphs vanish for non-zero external momenta, it yields a finite contribution for vanishing external momenta, which is the anomaly. Specifically, the integrated versions for the star gauge invariant axial anomaly in two and four dimensions were found to be,
\be
 \int d^2 x (\p_\mu j^{\mu(5)})= \int d^2 x ({\hat{\cal A}})=({1 \over{2\pi}}) \e^{\mu\nu} \int d^2 x  \, F_{\mu\nu} 
\label{14}
\ee
and,
\be
 \int d^4 x (\p_\mu j^{\mu(5)})= \int d^4 x({\hat{\cal A}}) =(- {1 \over{16\pi^2}})\e^{\mu\nu\rho\sigma} \int d^4 x  \,( F_{\mu\nu} *  F_{\rho\sigma})
\label{15}
\ee

The integrated expressions are in fact identical to the integrated versions of the star gauge covariant anomaly.  The origin of this  equivalence is contained in the fact that the two currents in  (\ref{1}), in their integrated forms, become identical, on using the commutativity of the star product within an integral. Indeed, by taking an actual point splitting regularisation and then taking their derivatives to get the anomalies leads to the same inference. This feature is implicit in the calculation of the two anomalies \cite{as, alt}.

  What about the unintegrated structure of the star gauge invariant anomaly?  In a recent collaborative work \cite{bg}, we found such an expression up to the first order in the noncommutative parameter $\theta$. The result, obtained by a point splitting regularisation, for arbitrary even dimensions, is given by,
\be
\p_\mu j^{\mu(5)}= N(1-\theta^{\alpha \beta}  \p_\alpha A_\beta)  (F\wedge F\wedge F .............\wedge F)_{n-fold} + O(\theta^2)
\label{16}
\ee
where the derivative acts on everything that is on its right, N is the normalisation given by,
\be
N={{2(-1)^{n+1}}\over {(4\pi)^n n!}}
\label{n}
\ee
and the wedge product is defined as usual. The derivative piece is in fact a typical non-planar contribution. To see that the above expression is star gauge invariant up to $O(\t)$, note that any gauge covariant object  $({\cal F}_{gc})$ transforming as (\ref{11}),  is converted into a gauge invariant object $({\cal F}_{gi})$ , up to $O(\t)$, by the following operation,
\be
{\cal F}_{gi} \rightarrow {\cal F}_{gc}-\t^{\a\b}\p_\a(A_\b{\cal F}_{gc}) +O(\t^2)
\label{gt}
\ee
as may be seen by a simple application of (\ref{11}) and (\ref{a}). Now $(*)(F\wedge F\wedge F .............\wedge F)$ transforms covariantly, where star operations are implied in all products. It differs from the corresponding expression with ordinary products by terms at least of $O(\t^2)$,
\be
(*)(F\wedge F\wedge F .............\wedge F)_{n-fold} = (F\wedge F\wedge F .............\wedge F)_{n-fold}+ O(\theta^2)
\label{wp}
\ee
This shows that the expression in (\ref{16}) is star gauge invariant up to the requisite order in $\t$.
It is easy to observe that the integrated expressions are compatible with (\ref{14}) and
(\ref{15}) because the extra contribution is a total boundary that simply drops out.
However it seems problematic to extend such perturbative computations to get the exact form of the star gauge invariant anomaly.  This is because the higher order corrections entail an infinite series of complicated diagrams so that their computation becomes highly formidable, if not impossible for all practical purposes.  Recourse has thus to be taken of some more general arguments.

\section{\bf{The anomaly from Ramond-Ramond couplings}}
 
Our proposal for the anomaly will satisfy the three  conditions:

\noindent  1. It should be star gauge invariant in the unintegrated form.

\noindent   2.  To first order in the noncommutative parameter, it should reproduce (\ref{16}).

\noindent   3. The integrated expressions should be identical to the integrated versions of the star gauge covariant anomaly, as for instance encoded in (\ref{14})  and  (\ref{15}). Furthermore since the integrated effect comes only from the zero momentum contribution, this should also be manifested.

   It is clear that any gauge invariant completion of  (\ref{16}) must involve the Wilson line operators.
These operators are path dependent and it is not clear how to attach them. Since the covariant anomaly is obtained as a generalisation of the ordinary axial anomaly, and since the covariant and invariant anomalies, in their integrated forms, are equivalent, it suggests that the origin of the invariant anomaly is also contained in the ordinary axial anomaly. The topological nature of the anomaly only strengthens this suggestion.  In other words, the star gauge invariant anomaly, under a certain map, should be identifiable with the ordinary anomaly.  This map is the well known Seiberg-Witten map, which transforms star gauge invariant expressions into ordinary gauge invariant expressions. The solutions to the map are not unique. They are defined modulo field redefinitions and gauge transformations \cite{ak}. String theory however provides us with a unique answer to the inverse map. Comparison between the Ramond-Ramond couplings in different  descriptions 
shows that the Fourier transform of the ordinary field strength maps to the following object in noncommutative space \cite{hl, oo, ms, lm},
\be
f_{\mu\nu}(k)=\int dx e^{i kx} ( \p_\mu a_\nu(x) -\p_\nu a_\mu(x))\rightarrow \int dx  *[e^{i kx} 
 \sqrt{det(1-\theta \bar F) }(\bar F (1-\theta \bar F)^{-1})_{\mu\nu} W(x, C)]
\label{18}
\ee
where *[ ... ] indicates that the star product has to be taken in the expansion inside the brackets,
W(x, C) is an open Wilson line chracterised by a straight path C,
\be
W(x, C)=P*exp(i\int_0^1 A_\mu(x+l\sigma) l^\mu d\sigma) ;\, \,  l^\mu= \theta^{ \mu\nu}k_\nu
\label{19}
\ee´
and $ \bar F$ is given in terms of the usual $F$ in noncommutative space by,
\be
 \bar F_{\mu\nu}= \int_0^1 F_{\mu\nu}(x+l\sigma)  d\sigma
\label{20}
\ee

Note that the same symbol $x$ is used to indicate the coordinates in the commutative and noncommutative descriptions related by the mapping (\ref{18}).

We  now propose the following structure for the star gauge invariant axial anomaly in a gauge theory defined in an arbitrary even $(d=2n)$ dimensional noncommutative space. The gauge invariant axial anomaly in usual space is taken and the field tensors occurring there are replaced by the corresponding objects provided by the map (\ref{18}) while the ordinary products are replaced by the star products. Thus the 2n-dimensional star gauge invariant anomaly is given by,
\be
\p_\mu j^{\mu(5)}= N* (f\wedge f\wedge f .............\wedge f)_{n-fold}
\label{21}
\ee
Here N is the standard normalisation and the use of the map (\ref{18}), translated in coordinate space, is implied  for all the $f's$.

 Since the above expression is star gauge invariant by construction, it trivially satisfies the first criterion mentioned earlier.   The other two conditions are next proved, one by one.
 
 \subsection{{\bf Computation up to first nontrivial order in $\t$}}
 
  To first order in the noncommutative parameter, the anomaly  must reproduce (\ref{16}).  A straightforward algebra leads to the following simplification in (\ref{18}),
\be
f_{\mu\nu}(x) \rightarrow F_{\mu\nu}(x) +\theta^{ \lambda\rho}  \Big (\p_\rho(A_ \lambda F_{\mu\nu})(x)
+{1\over 2} F_{\mu\nu}(x) F_{\lambda\rho}(x)  -F_{\mu\lambda}(x) F_{\nu\rho}(x)  \Big )+O(\t^2)
\label{22}
\ee
Using the definition (\ref{10}) for the field tensor, the above mapping reduces to,
\be
f_{\mu\nu}(x) \rightarrow F_{\mu\nu}(x) +\theta^{ \lambda\rho}  \Big (A_ \lambda \p_\rho F_{\mu\nu}(x)
  -F_{\mu\lambda}(x) F_{\nu\rho}(x)  \Big ) +O(\theta^2)
\label{23}
\ee
Noting further that upto $O(\theta)$, the star products in (\ref{21}) can be replaced by the ordinary products (see (\ref{wp}), we get,
\be
\p_\mu j^{\mu(5)}= N[( F + n(F\theta F)) \wedge (F\wedge F\wedge F.............\wedge F)
+\theta^{ \lambda\rho}  \Big (A_ \lambda \p_\rho (F\wedge F\wedge F.............\wedge F)\Big )]
\label{24}
\ee
where the matrix notation has been introduced,
\be
(F\theta F)_{\lambda \rho}= F_{\lambda \mu}\theta^{ \mu\nu}  F_{\nu \rho}
\label{25}
\ee
The derivative in the second term is pulled out so that it acts on the $A_ \lambda$. This will generate an extra piece which, up to the order we are  interested, is proportional to the field tensor so that, 
\ber
\p_\mu j^{\mu(5)} &=& N[( F + n(F\theta F)) \wedge (F\wedge F\wedge F.............\wedge F) \cr \ \
&+&\theta^{ \lambda\rho}  \p_\rho  \Big (A_ \lambda (F\wedge F.............\wedge F)\Big )
+{1\over 2}\theta^{ \lambda\rho}F_ { \lambda\rho}(F\wedge F.............\wedge F)]
\label{26}
\eer
Using the identity,
\be
\theta^{ \lambda\rho}F_ { \lambda\rho}(F\wedge F.............\wedge F)=-2n (F\theta F) \wedge (F\wedge F\wedge F.............\wedge F)
\label{27}
\ee
the final result turns out to be,
\be
\p_\mu j^{\mu(5)}= N(1-\theta^{\alpha \beta}  \p_\alpha A_\beta)  (F\wedge F\wedge F .............\wedge F)_{n-fold} + O(\theta^2)
\label{28}
\ee
 which exactly reproduces  (\ref{16}).  As in (\ref{16}), the derivative acts on everything that is on the right. Thus the second condition is also satisfied.
 
 \subsection{{\bf The integrated anomaly}}

   Finally, the third condition is considered. This analysis is separated in two parts. First, the two dimensional case is discussed, after which the higher dimensions will be considered.
 
 \bigskip  
   
   \noindent{\it{The two dimensional analysis}}:
   
   \bigskip
   
   The star gauge invariant anomaly now becomes,
\be
 \p_\mu j^{\mu(5)}= {1 \over{2\pi}}   \e^{\mu\nu}f_{\mu\nu} 
\label{29}
\ee
where, as usual, $f_{\mu\nu}$ has to be replaced by the r.h.s. of the map  (\ref{18}). In considering the integrated version of the above equation, it is crucial to note that there are two types of contributions coming from  (\ref{18}): one of these are total derivative terms which drop out on integration while the other, nonderivative terms survive. These terms are just given by the factor (after setting $k=0$) multiplying the open Wilson line. Thus,
\be
 \int d^2 x (\p_\mu j^{\mu(5)})={1 \over{2\pi}}  \e^{\mu\nu}\int d^2 x (*)  \sqrt{det(1-\theta F) } (F (1-\theta F)^{-1})_{\mu\nu}
\label{30}
\ee
where star products are always implied in all the expansions. It becomes convenient to multiply both sides of this equation by $\theta$ to exploit  the identification $ \theta^{\alpha \beta}=\theta \e^{\alpha \beta}$. It yields,
\be
\theta \int d^2 x (\p_\mu j^{\mu(5)})=-{1 \over{2\pi}}\int d^2 x (*)  \sqrt{det(1-\theta F) }
tr[\theta F (1-\theta F)^{-1})]
\label{31}
\ee
The two factors in the integrand can be simplified to yield,
\be
 \sqrt{det(1-\theta F) }= exp{1 \over 2} [tr log(1-\theta F)]=exp[ log(1+\theta F_{12})]=1+\theta F_{12}
\label{32}
\ee
and,
\be
tr[\theta F (1-\theta F)^{-1})]= -\theta  \e^{\mu\nu} F_{\mu\nu} (1+\theta F_{12})^{-1}
\label{33}
\ee
obtained by a straightforward expansion of the determinant and the rational fraction. Using the associativity of the star product within an integral, we finally obtain,
\be
 \int d^2 x (\p_\mu j^{\mu(5)})={1 \over{2\pi}} \int d^2 x   (\e^{\mu\nu} F_{\mu\nu})
\label{34}
\ee
which reproduces the standard expression for the integrated version of the two dimensional star gauge invariant  anomaly. Observe that the entire dependence on the noncommutative parameter is contained in the definition of the field tensor.

\bigskip

\noindent{{\it The higher dimensional analysis}}:

\bigskip

The analysis is next extended to four and higher dimensions. For these dimensions it is not possible to use the simplification for the noncommutative parameter that was done in the two dimensional example. Before proceeding further we note that the integrated version of the field tensor following from (\ref{18}) corresponds to the zero momentum limit $(\int dx  f_{\mu\nu}(x)= f_{\mu\nu}(k=0))$. This means that the contribution to the integrated anomaly is given by the zero external momentum part, which fits in with the general discussion done earlier and also mentioned in point 3. 

For higher dimensions, instead of directly taking the  star products of the field tensors 
occurring in (\ref{21}), and then taking the integral {\footnote{We shall also discuss this approach subsequently.}} , there is a more elegant way of proceeding.
 The point is that the mapping (\ref{18}) is only one among a series of mappings obtained by a study of the Ramond-Ramond couplings in different descriptions. Since they have a common origin, it is possible to infer that same results would be obtained from these identities, as would be obtained by a direct application of the inverse Seiberg -Witten map (\ref{18}).

The various identities are obtained by a study of the Wess-Zumino action which,  in the noncommutative description,  is expressed in terms of  differential 2n-forms, $Q_{2n}$, given by,
\ber
Q_{2n}(k)= \int dx  *[e^{i kx} 
 \sqrt{det(1-\theta \bar F) }\Big( (\bar F (1-\theta \bar F)^{-1}) &\wedge& (\bar F (1-\theta \bar F)^{-1})
................ \cr
& \wedge& (\bar F (1-\theta \bar F)^{-1})\Big)
 W(x, C)]
\label{35}
\eer
The two form is special since it can be identified with the corresponding two form in the commutative case leading to the inverse Seiberg-Witten map  (\ref{18}). A similar statement for higher forms is not possible due to the presence of derivative corrections. We shall return to this point later on.  However it can be inferred that since  $\int dx Q_{2n}(x)$ yields the topological charge,  it is possible to identify it with the corresponding topological charge in the commutative description, since such an object is an invariant that cannot depend on how it is calculated. Thus the following mapping is obtained \cite{lm},
\ber
 \int dx (f\wedge f)&=& Q_{4}(k=0)\cr
  \int dx (f\wedge f\wedge f)&=& Q_{6}(k=0)\cr
  . &=& .\cr
   . &=& .\cr
    . &=& .\cr
    . &=& .\cr
 \int dx (f \wedge f\wedge ..........\wedge f)&=&   Q_{2n}(k=0)
\label{36}
\eer
where,
\ber
 Q_{2n}(k=0)&=&  \int dx Q_{2n}(x)=\int dx  *[\sqrt{det(1-\theta  F) }\Big(  (F (1-\theta  F)^{-1}) \wedge ( F (1-\theta
F)^{-1})
\cr
 .... &\wedge&(F (1-\theta  F)^{-1})\Big)]
\label{37}
\eer
According to our proposal therefore the integrated version of the 2n-dimensional star gauge invariant anomaly is  just given by the above formula; i.e.,
\be
 \int d^{2n} x (\p_\mu j^{\mu(5)})=
N \int d^{2n}x  *[\sqrt{det(1-\theta  F) }\Big(  (F (1-\theta  F)^{-1}) \wedge ( F (1-\theta
F)^{-1})
 \wedge ....... (F (1-\theta  F)^{-1})\Big)]
\label{38}
\ee
where N is the normalisation of the anomaly. Since the individual terms transform covariantly, it is clear that the integrated version is star gauge invariant. However this is not the desired form (see, for instance, (\ref{15}). It is possible to simplify this expression considerably to get this form. We first do the proof for constant fields (i.e. where the derivatives on F can 
be ignored) and then argue that the result holds in general. For constant fields the star product gets replaced by the ordinary product. The integrand is thus given by,
\be
A=
\sqrt{det(1-\theta  F) }\Big(  {\cal F}  \wedge  {\cal F} 
 \wedge .......{\cal F} 
\Big)
\label{39}
\ee
where,
\be
  {\cal F}  = (F (1-\theta  F)^{-1})
\label{40}
\ee
and all products are ordinary ones. The $\theta$ dependence in the above expression (\ref{39})
comes from the explicit structure in the determinant, in $ {\cal F}$ , as well as implicitly from the definition of $F$. It is now shown that the explicit dependence actually  
vanishes.  Taking the derivatives with respect to the explicit dependence, we get,
\be
\d{\cal F}_{\mu\nu}= \d\t_{\a\b}{{\p  {\cal F}_{\mu\nu}}\over {\p\theta_{\alpha\beta}}}  =  \d\t^{\a\b}
\Big({\cal F}_{\mu\alpha}{\cal F}_{\beta\nu}
-{\cal F}_{\mu\beta}{\cal F}_{\alpha\nu}\Big)= 2({\cal F}\d\t{\cal F})_{\mu\nu}
\label{41}
\ee
and similarly,
\be
\d \Big(\sqrt{det(1-\theta  F) } \Big) = {\d\theta^{\alpha\beta}}\Big( \sqrt{det(1-\theta  F) } \Big){\cal F}_{\alpha\beta}\label{42}
\ee
Using these relations, the required variation in  (\ref{39}) is expressed as,
\be
\d A=
\sqrt{det(1-\theta  F) }\Big( (\d\theta^{\alpha\beta})  {\cal F} _{\alpha\beta} +2n ({\cal F}(\d\theta){\cal F})
\wedge\Big)
\Big(  {\cal F} 
 \wedge .......{\cal F} 
\Big)
\label{43}
\ee
On exploiting the identity (\ref{27}), valid for any antisymmetric tensors, the expression on the right side of the above equation vanishes. This means that there is no explicit dependence on 
$\theta$,
 so that  (\ref{39}) is simplified by setting $\theta=0$, wherever it occurs explicitly. Thus,
\be
\sqrt{det(1-\theta  F) }\Big(  {\cal F}  \wedge  {\cal F} 
 \wedge .......{\cal F} 
\Big) = \Big(  F \wedge F
 \wedge ....... F \Big)
\label{39a}
\ee

The final result for the integrated star gauge invariant anomaly, in the constant field approximation, is thus given by,
\be
 \int d^{2n} x (\p_\mu j^{\mu(5)})=
N \int d^{2n}x \Big(  F \wedge F
 \wedge ....... F \Big)
\label{44}
\ee
To extend this result for general field tensors, we recall that the expression must be star gauge invariant. Furthermore, the star products that were discarded must be reintroduced. The only (and also the most natural) possibility compatible with (\ref{44}) is to replace the ordinary products by the star products. The integrated anomaly in the general case then becomes, 
\be
 \int d^{2n} x (\p_\mu j^{\mu(5)})=
N \int d^{2n}x (*)\Big(  F \wedge F
 \wedge ....... F \Big)
\label{45}
\ee
where all products are now replaced by the star products.
Since the $F's$ transform covariantly,  the integrated anomaly is invariant due to the cyclicity of the star product within the integral. Any other modification in (\ref{45}), that preserves gauge invariance,  would violate (\ref{44}). For instance if one thinks of a new term where the field tensors are replaced by their covariant derivatives (which also transform covariantly), that term actually gets reexpressed in terms of   (\ref{44}),  As an example in four dimensions, consider,
\be
 \e^{\mu\nu\rho\sigma}(*)D_\mu D_\nu F_{\rho\sigma}= {1\over 2}\e^{\mu\nu\rho\sigma}(*)[D_\mu, D_\nu] F_{\rho\sigma}= {i\over 2}\e^{\mu\nu\rho\sigma}(*)F_{\mu\nu} F_{\rho\sigma}
\label{47}
\ee
which, in the constant field approximation, would alter  (\ref{44}).

We have thus shown that the last condition (3), mentioned earlier, is satisfied.

   Expectedly, the same conclusion ((\ref{44}) and hence (\ref{45})) can also be reached by directly starting from the inverse Seiberg-Witten map (\ref{18}) and using our proposal for the unintegrated anomaly. For constant field approximation, the mapping is simply given by \cite{sw},
\be
f \rightarrow {\cal F} 
\label{48}
\ee
Hence the invariant anomaly, in its unintegrated form, is given by,
\be
\p_\mu j^{\mu(5)}=
N\Big( {\cal F} 
 \wedge {\cal F} 
 \wedge .......{\cal F} \Big)
\label{49}
\ee
Using (\ref{39a}), the above relation is written as,
\be
\p_\mu j^{\mu(5)}={N\over\sqrt{det(1-\theta  F)} }\Big(  F \wedge F
 \wedge ....... F \Big)
\label{50}
\ee
In the constant field approximation, the product of terms in the expansion of the determinant  and the wedge chain can be expressed in terms of total derivatives, so that,
\be
\p_\mu j^{\mu(5)}=N \Big(  F \wedge F
 \wedge ....... F \Big)\,\,\, + total \,\,\, derivative \,\,\, terms
\label{51}
\ee
These are the type of correction terms that prevent an exact mapping of the higher (than 2) forms (\ref{35}) in the commutative and noncommutative descriptions. Such terms also occur in the comparison of the Born-Infeld  lagrangians in the two pictures \cite{sw}. Indeed it is the Born-Infeld action that remains invariant. Likewise, we find here that the integrated version of (\ref {51}) reproduces (\ref {44}), thereby completing the proof.

As a simple corollary of (\ref{50}), it is possible to relate the star gauge invariant and covariant anomalies in the constant field approximation. The point is that the factor multiplying the determinant in (\ref{50}) is just the star gauge covariant anomaly (\ref{9}). Hence we obtain,
\be
\p_\mu j^{\mu(5)}=  {1\over\sqrt{det(1-\theta  F)} } D_\mu J^{\mu(5)}
\label{cov}
\ee

\bigskip

\section{\bf{Seiberg-Witten transformation and integrated anomalies: an alternative derivation; Elliott formula}}

\bigskip

It is also possible to provide an alternative  evidence that (\ref{45}), under the Seiberg-Witten transformation, is equivalent to the integrated anomaly in the usual space.  Contrary to the previous approach, the equivalence will be established by directly working with general fields, avoiding the passage through the constant field approximation. Of course for two dimensions the result was already established (see (\ref{34})), directly in general terms, based on the simplification of the noncommutativity parameter. 

Basically, the idea is to  invert the process adopted till now, i.e.; instead of using the inverse map , the usual map will have to be used. However, unlike the inverse map, the map from the noncommutative to the commutative picture is not known in a closed form.  But that is not important because the basic correspondence between the noncommutative and commutative descriptions is known to be given by \cite{sw},
\be
\d F_{\mu\nu}(\t)={1\over 4}\d\t^{\rho\sigma}\Big[2\{F_{\mu\rho}, F_{\nu\sigma}\}_+
-\{A_\rho , D_\sigma F_{\mu\nu} +\p_\sigma F_{\mu\nu} \}_+\Big]
\label{52}
\ee
subjected to the initial condition that for vanishing $\theta$, the noncommutative field tensor reduces to the usual one $(F_{\mu\nu}(\t =0)=f_{\mu\nu})$. The covariant derivative has been defined in (\ref{8}) and the following notation has been introduced,
\be
\{ A, B\}_+= A*B +B*A
\label{53}
\ee
The map (\ref{18}) is just a solution to this equation.  
Let us now study the response of this transformation on the integrated anomalies. As before, the analysis is split into two parts. First, the two dimensional case is considered.

\newpage

\noindent{\it{ The two dimensional example}}:

\bigskip

Under the transformation (\ref{52}), the anomaly (\ref{14}) responds as,
\be
\int \, d^2x (\d \hat A) ={\e^{\mu\nu}\over {2\pi}}\int\,d^2x \, (*)\d\t^{\rho\sigma}\Big(F_{\mu\rho} F_{\nu\sigma}-A_\rho
\p_\sigma F_{\mu\nu} - iA_\rho A_\sigma F_{\mu\nu}\Big)
\label{1.2}
\ee 
where repeated use has been made of the  the associativity of the star product within an integral,
\be
\int \Big(A*B*C\Big)=\int \Big(B*C*A\Big)=\int \Big(C*A*B\Big)
\label{55}
\ee
Up to a divergence term that is discarded, it can be written as,
\ber
\int \, d^2x\,(\d \hat A) &=&{\e^{\mu\nu}\over {2\pi}}\int\,d^2x\, (*)\d\t^{\rho\sigma}\Big(F_{\mu\rho} F_{\nu\sigma}+{1\over 2}
 F_{\mu\nu}F_{\sigma\rho}\Big)\cr
 &=& 0
\label{2.2}
\eer 
with the last line following identically from two dimensional properties. This means there is no $\t$-dependence and the complete expression is equivalent to its initial value obtained by setting $\t=0$. Thus the integrated versions of the invariant anomalies in the two descriptions get mapped, so that,
\be
{\e^{\mu\nu}\over {2\pi}}\int\, d^2x F_{\mu\nu}\rightarrow {\e^{\mu\nu}\over {2\pi}}\int\, d^2x f_{\mu\nu}
\label{3.2}
\ee
which is what we set out to prove, reconfirming our assertion that the noncommutative anomaly is obtained from the usual anomaly through the use of the Seiberg-Witten transformation. Since there is only one component of the field tensor, it can also be expressed as,
\be
\int\,d^2x F \rightarrow \int\,d^2x  f
\label{4.2}\ee

We next discuss higher dimensions. For four dimensions, the result analogous to (\ref{3.2}) is explicitly done which is extended to other higher dimensions by a topological argument.

\bigskip

\noindent{\it{Four dimensional analysis}}:

\bigskip

The response of 
 (\ref{52}) on the four dimensional integrated anomaly (\ref{15}) is given by,
\be
  \int d^4 x  (\d{\hat{\cal A}}) =- {1 \over{8\pi^2}} \e^{\mu\nu\rho\sigma} \int d^4 x  (\d F_{\mu\nu})* F_{\rho\sigma}
\label{54}
\ee
Using standard manipulations with the star products, we find,
\be
  \int d^4 x  (\d{\hat{\cal A}}) ={1 \over{32\pi^2}}  \e^{\mu\nu\rho\sigma} \int d^4 x (*) 
  \Big(4(F\d\t F)_{\mu\nu} F_{\rho\sigma}+\d\t^{\a\b}\Big(2A_\a \p_\b(F_{\mu\nu}F_{\rho\sigma})
  +i[A_\a, A_\b] F_{\mu\nu}F_{\rho\sigma}
  \Big)
\label{56}
\ee
where star operations are implied in all products and the notation of (\ref{25}) has been used. 
Up to a boundary term that is discarded, we obtain,
\ber
  \int d^4 x  (\d{\hat{\cal A}})& =&{1 \over{32\pi^2}}  \e^{\mu\nu\rho\sigma} \int d^4 x (*) 
  \Big(4(F\d\t F)_{\mu\nu} F_{\rho\sigma}+\d\t^{\a\b}F_{\a\b}F_{\mu\nu}F_{\rho\sigma}\Big)\cr
  &=& 0
\label{57}
\eer
where the last equality is a consequence of
 an identity, valid for any pair of second rank antisymmetric tensors,
\be
\int \, d^4x\, (*)\Big(4((ABA) \wedge A) +B^{\a\b}A_{\a\b}(A\wedge A)\Big) = 0
\label{58}
\ee
It is the $n=2$ result of the general identity (\ref{27}); however it is only valid within an integral if the ordinary products are replaced by the star products. This means there is no $\t$-dependence and the integrated versions of the invariant anomalies in the two descriptions get mapped, so that,
\be
-{1\over {16\pi^2}} \int\, d^4x\, (*) (F\wedge F) \rightarrow\, -{1\over {16\pi^2}} \int\, d^4x\, (f\wedge f)
\label{59}
\ee
This concludes the analysis for four dimensions.

It is possible to proceed with other higher dimensions in the same way. However , we note that (\ref{4.2}) and (\ref{59}) yield the topological charges in two and four dimensions in the commutative description. Thus the objects to which these are mapped must represent the topological charges in the noncommutative description. Since for arbitrary dimensions the charges are just given by the wedge extensions in the commutative picture, these should be given by the star wedge extensions in the other picture. Thus the integrated anomalies in the two versions must be mapped by the Seiberg-Witten transformation, so that,
\be
\int\, d^{2n}x (*)(F\wedge F\wedge F .....\wedge F)\rightarrow\int\,d^{2n}x(f\wedge f\wedge f.....\wedge f)
\label{68a}
\ee

There is a subtle point that is next discussed. We have seen that the noncommutative Chern character,
\be
 {\cal C}_{nc}=Tr(*) (e^F)= \int\, (*)
e^F
\label{64}
\ee
is a straightforward generalisation of the commutative version $(\int e^f)$.  But it is known that, in the general case, the noncommutative Chern character is given by the Elliott formula \cite{e, 
hl},
\ber
{\cal C}_E&=& Tr(*) [\sqrt{det(1-\theta  F) } e^{ (F (1-\theta  F)^{-1})}] \cr
            &=&\int dx (*) [\sqrt{det(1-\theta  F) } e^{ (F (1-\theta  F)^{-1})}] 
\label{68b}
\eer
where, for the $U(1)$ case, there is no trace over the group indices. Actually the same formula is also obtained in string theory by a comparison of the Ramond-Ramond couplings  \cite{lm}. It is given here by (\ref{37}). But it was shown there the complete dependence on the $\t$ parameter is contained only in the implicit dependence through the definition of the field tensor.  Thus, setting the explicit $\t =0$, (\ref{68b}) reduces to (\ref{64}), thereby simplifying the Elliott formula,
\be
{\cal C}_E = {\cal C}_{nc}
\label{el}
\ee

The Chern character is known to be both invariant as well as closed. It can therefore be expressed in terms of Chern Simons forms $\Omega$ such that,
\be
\int  (*)(F\wedge F\wedge F .....\wedge F)=\int d\Omega_{2n+1}
\label{cs}
\ee
with,
\be
\Omega_{2n+1}=n\int_0^1dt (*)\Big(A\wedge F_t\wedge F_t............\wedge F_t\Big)
\label{cs1}
\ee
where $F_t=t dA+i t^2 A*A$. Since the integrated anomalies are mapped by the Seiberg-Witten transformation, it is expected that the noncommutative Chern-Simons action would likewise be mapped to the usual Chern-Simons action.  This has been explicitly checked in the three dimensional (n=1) case  in \cite{gs} {\footnote {Note, however, that there are subtleties in this mapping \cite{p} for topologically nontrivial configurations.}}. Also, the structure of the three dimensional Chern-Simons action in noncommutative space, found \cite{gs} by an explicit loop calculation, agrees with (\ref{cs1}). 

\bigskip

\section{{Discussions}}

\bigskip

We have proposed an unintegrated form of the  star gauge invariant axial anomaly in a noncommutative gauge theory. It was based on the Seiberg-Witten transformation relating the commutative (usual) and noncomutative descriptions. The precise solutions \cite{hl, oo, ms, lm} that were used were found earlier by string theory calculations based on a study of Ramond ramond couplings in different descriptions. The proposal for the anomaly was shown to be compatible with existing results in the literature. 

The explicit computation of the gauge invariant axial anomaly is known to be complicated by the appearance of non-planar graphs which, by naive considerations, are expected to vanish.   However as discussed in \cite{as, bg,alt, n},  these graphs suffer from the problem of UV/IR mixing. 
A proper treatment shows, that for the integrated version of the invariant anomaly, there is a contribution only from the zero external momentum \cite{as, alt}. This was clearly shown in our analysis. In fact the unintegrated result was obtained by smearing the zero momentum result to an open Wilson line, such that it was a solution of the Seiberg-Witten transformation.

We have clarified several aspects of the integrated anomalies and their relation to the noncommutative Chern character. It is generally thought that the noncommutative Chern character is just a simple star deformation of the usual Chern character. In general it is not true, rather it is given by the Elliott formula \cite{e}, which is also confirmed by string theory calculations 
involving D-branes \cite{lm}.  For trivial topologies a meaningful separation of the trace in the Elliott formula (\ref{68b}), as an integration  over space and a trace over the group indices, is possible. It was also shown that the complicated topological structure simplified  into the standard star deformation of the commutative Chern character. Expectedly, the Chern-Simons forms followed from this character.

In the constant field approximation, the unintegrated invariant and covariant anomalies were shown to be proportional, with a field dependent normalisation (\ref{cov}). It was reassuring to note that the integrated versions became identical, both yielding the expected Chern character.

A consequence of this analysis was that the integrated noncommutative anomalies were mapped to the corresponding usual integrated anomalies by the Seiberg-Witten equation. This was shown in two distinct ways; once by starting from the usual axial anomaly and using the solutions of the 
equation, then by taking the noncommutative version and using the basic transformation itself.
Since the solutions of the map are not unique \cite{ak}, it appears that those found naturally from string theory might have a special status {\footnote {Such a possibility was raised in footnote 4 of \cite{oo}.}},  since it correctly led to  the integrated noncommutative anomalies from the usual ones.

It is also possible to obtain the chiral anomalies, in analogue of what was done for the usual case \cite{rb}. Consider, for instance a chiral theory, where the Dirac operator in (\ref{0}) is replaced as,
\be
D_\mu\s=\p_\mu\s +i P_+A_\mu*\s; \,\,\, P_+={{1+\gamma_5}\over 2}
\label{d1}
\ee
while the classically conserved chiral gauge invariant current is given by, 
\be
 j^{\mu(R)}=\bs *\g^\mu P_+\s
 \label{d2}
 \ee
 Taking the vacuum expectation value, we find,
 \be
 < j^{\mu(R)}> = Tr\Big(\g^\mu P_+(*) <X|(\g^\mu(P_\mu +P_+ A_\mu))^{-1}|X>\Big)
 \label{d3}
 \ee
 where the expression has been written in the $(X, P)$ basis and star products are implied in the expansion of the inverse propagator. Noting that $P_+$ is a projection operator, this simplifies to,
 \be
 < j^{\mu(R)}> = Tr\Big(\g^\mu P_+(*) <X|(\g^\mu(P_\mu + A_\mu))^{-1}|X>\Big)
 \label{d4}
 \ee
 implying that the chiral current can be interpreted as comprising two pieces, a vector and an axial vector current, both defined  in a vector gauge theory. Using a vector gauge invariant regularisation, the anomaly in the first part is zero. The other part is just half of the axial anomaly, whose form has been proposed here. Thus the chiral anomaly is obtained.
 
 There are several directions in which the present work may be extended. One possibility is to study the structure of anomalous commutators in noncommutative theories. These are connected to divergence anomalies by a set of consistency relations \cite{b}, which were exploited in the usual case \cite{bg1}. A knowledge of the unintegrated anomalies is essential which has been provided here. Also, the role of the Seiberg-Witten transformation in the study of commutator anomalies is relevant. Our proposal for the gauge invariant anomaly could be tested for spaces other than ${\bf R^{2n}}$, as for instance in the case of the fuzzy sphere. The usual problems of calculating the unintegrated star gauge invariant anomaly persist there and recourse is taken to the gauge covariant anomaly \cite{asn}.  Our work can have implications in formulating the standard model on noncommutative spaces \cite{w}.  Such models are generally studied by converting the original noncommutative theory to an effective theory in the usual space, by using the Seiberg-Witten map. The fact that the anomalies get identified in the different descriptions, through this map, supports the construction of such an effective theory. 
  
 \bigskip
 
 {\Large{\bf{Acknowledgements}}}: I thank the Japan Society for Promotion of Science for support , Izumi Tsutsui for discussions and the members of the theory group, KEK, for their gracious hospitality.

\end{document}